\documentstyle[12pt,aaspp4]{article}

\begin{document}

\title{Gamma-Ray Burst Afterglows with Energy Injection:
Homogeneous Versus Wind External Media$^0$}  
\author{ WANG Wei$^{1,2}$, DAI Zi-Gao$^{1}$}

\affil{$^1$Department of Astronomy, Nanjing University, Nanjing 210093 }
\affil{$^2$Beijing Astronomical Observatory, National Astronomical
Observatories, Chinese Academy of Sciences, Beijing 100012 } 
\affil{({\it Received 6 December, 2000})}

\altaffiltext{0}{Supported by the National Natural Science
Foundation
of China under grant No.19825109 and the National 973 project.}
 
 {\sl Assuming an adiabatic evolution of a gamma-ray burst (GRB)
fireball interacting with an external medium,  we calculate the hydrodynamics
of the fireball with energy injection from a strongly magnetic millisecond
pulsar through  magnetic dipole radiation, and obtain the light curve of
the optical afterglow from the fireball by synchrotron radiation. Results
are given both for a homogeneous external medium and for a wind  ejected by
GRB progenitor. Our calculations are also available in both ultra-relativistic
and non-relativistic phases. Furthermore, the observed R-band light curve of
GRB{000301C} can be well fitted in our model, which might provide a probe of
the properties of GRB progenitors.} 
 

Since the measurement of the redshift of GRB{970508}, gamma-ray bursts(GRBs) 
as the cosmological origin are confirmed.$^{[1]}$ 
Thus, one of the most important issues is the nature of the central engine. 
The popular theoretical explanation of their radiation 
properties has been commonly thought to be the fireball model$^{[2]}$, 
in which GRBs result from the dissipation of the kinetic energy of the 
relativistically expanding fireballs. Two main ideas 
have been proposed to realize the dissipation: the internal and external shock model.
In present letter, we will consider the external shock model which reproduces 
very well the delayed 
emission at lower energy. 

A number of afterglows from GRBs have been observed at X-ray, optical and radio 
wavelengths, whose major features can be 
explained by simple standard models.$^{[3,4]}$ We consider a newborn
millisecond  pulsar with a strong magnetic field at the center of 
the fireball,$^{[5]}$ which may be formed by several models: accretion-induced 
collapse of magnetized white dwarfs,$^{[6]}$ merge of 
two neutron stars,$^{[7]}$ and accretion-induced phase transitions 
of neutron stars to strange stars.$^{[8,9]}$ It has been proposed that very 
strongly magnetized pulsars 
may be the central engine of GRBs.$^{[5]}$ It is natural to expect that if a
GRB results  from the birth of a strongly magnetic millisecond pulsar, then
after the main GRB,  the pulsar continuously supplies energy to 
the fireball through magnetic dipole radiation. And during the process, we
haven't consider gravitational radiation due to the very high
surface temperature of the newborn pulsar.$^{[10]}$ The power of the pulsar is
radiated away mainly through electromagnetic waves  with frequency of
$\omega=2\pi/P$, where $P$ is the period of the pulsar. Once the
electromagnetic waves propagate in the shocked interstellar medium (ISM), they
will be absorbed for $\omega_p>\omega$,$^{[5]}$  where $\omega_p$ is the
plasma frequency of the shocked ISM. 

At the center of the fireball, the pulsar loses its rotational energy through 
magnetic dipole radiation,$^{[7]}$ 
whose power is given by
\begin{equation}
L=4\times 10^{43} B_{\perp,12}^2 P_{i,ms}^{-4} R_{NS,6}^6 
      {\left(1+{t_b \over T}\right)}^{-2}{\rm erg~s^{-1}}, 
\end{equation}
where $B_{\perp,12}=B_s {\sin\theta}/{10^{12}}$in units of G, $B_s$ is the
surface dipole field strength, $\theta$ the angle between the magnetic
dipole moment and rotational axis, $P_{i,ms}$ the initial period in units
of 1ms, and $R_{NS,6}$ is the neutron star(NS) radius in units of $10^6$cm,
$t_b$ is the burster time and $T$ is the initial  spin-down timescale defined
by $T=P/{2\dot{P}}$.

In the following we will consider the adiabatic expansion of a fireball. In the fixed frame, 
the total kinetic energy 
of the fireball is$^{[11]}$
\begin{equation}
E_k=(\gamma-1)(M_0+M_{sw})c^2+\gamma U,
\end{equation}
where $\gamma$ is the bulk Lorentz factor, $M_0$
is the rest mass ejected from the GRB central engine, $M_{sw}$ is the rest mass of 
the swept-up medium, $U$ is the internal energy. As 
Huang {et al}.$^{[12,13]}$ suggested $U=(\gamma-1)M_{sw}c^2$,
available in both ultra-relativistic and non-relativistic phases. 
For an adiabatic fireball, the change of the kinetic energy equals to the energy 
which the fireball has obtained from the pulsar through magnetic dipole radiation:
\begin{equation}
\frac {d E_k}{d t_b}=(1-\beta)L(t),
\end{equation}
where $\beta=(1-1/\gamma^2)^{1\over2}$, $t=t_b-R/c$ is the 
observed time, and $R$ the blast wave radius.
From Eqs (2) and (3), we can find
\begin{equation}
\frac {d \gamma}{d t_b}=\frac {(1-\beta)L(t)}{(M_0+2\gamma M_{sw})c^2}
      -\frac{\gamma^2-1}{M_0+2\gamma M_{sw}}\frac{d M_{sw}}{d t_b}.
\end{equation}

Generally, it is assumed that the medium surrounding the GRB source is homogeneous, $n(R)=n_{*}$.
However, if a collapsing massive star$^{[14]}$ is the origin of a  relativistic
fireball, 
the circum-burst medium is the wind ejected by the star prior to
its collapse, whose density decreases outwards. Recently, the discovery of
the connection between supernovae and GRBs is a strong support to the
assumption that 
 some GRBs come from collapsars.$^{[15]}$ 
The dynamical
evolution and afterglow in the wind model 
are expected to be very different
from those in the homogeneous case. Significant work in this direction has
been done 
by some researchers.$^{[3,16]}$
 The baryon number density of the
wind medium
 is expected to be $n(R)=AR^{-2}$,
$A$ is scaled to $A=3.0\times{10^{35}}A_{*}{\rm cm^{-1}}$,$^{[16]}$
and $A_* =
\dot{M}/v$ where $\dot{M}$ is the mass loss rate of massive star in unit of
${\rm M_\odot yr^{-1}}$ that has ejected the wind at constant speed $v$ in
units of ${\rm 10^3 km s^{-1}}$. In present work, we give more detailed
calculations in both the homogeneous external medium and the wind, and
investigate the difference between the dynamical evolution and light curves of
afterglows arising  for two types of external medium to find ways for
distinguishing the two models.

Based on the model described above, we have evaluated the 
propagation of the blast wave numerically by taking
$E_0=10^{51}{\rm ergs}$, $n_{*}=1{\rm cm^{-3}}$, $A_{*}=1,$
$M_0=5\times{10^{-6}}{\rm M_{\odot}}$, $P_i=1{\rm ms}$,
$R_{NS}=10^6{\rm cm}$, and $I=2.0\times{10^{45}}{\rm g~cm^2}$.
To describe completely the properties of the case with energy injection, we
also calculate afterglows without energy injection as a comparison. 
In Figs. 1 and 2, the solid lines correspond to the case 
without energy injection, the dotted lines to the surface 
magnetic field strength of the pulsar, $B_s=10^{12}{\rm G}$, the dashed lines 
to $B_s=10^{13}{\rm G}$ and the dash-dotted lines to $B_s=10^{14}{\rm G}$.

Figure 1 shows the evolution of the bulk Lorentz factor in the homogeneous 
ISM and  wind models. From this figure, 
we can see that the bulk Lorentz factor decreases faster          
firstly but more slowly later in the wind model than in the homogeneous case.
Comparing the cases of with and without energy
injection, we find that the bulk Lorentz factor in the former is slightly
larger than that of the latter. With different surface magnetic field
strengths of the pulsars, the timescales at which the effect of energy
injection on the evolution of the fireball becomes significant vary
accordingly.

To compare with observations, we need calculate synchrotron radiation 
from the shocked medium, in which
we take the same electron distribution as in Dai {et al.}.$^{[17]}$ In Fig.
2, we show the computed R-band optical light curves.  In these figures, some
of the parameters are fixed ($p=2.5,\xi_e=0.1,\xi_B=0.01,D_L=3{\rm Gpc}$).
Figure 2a corresponds to  the homogeneous case, and Figure 2b to the wind
case. We clearly see that the afterglow light curves in the wind model  are
steeper than those in the homogenous medium, verifying our expectation.
Furthermore, we find that the magnitude of the optical afterglow from the
fireball first decreases rapidly with time, subsequently flattens, and finally
declines again. From the existence of the flattening in the light curves, we
may successfully explain 
the flattening and steepening optical features
observed in some GRB afterglows 
which contradict the results of the standard
model. 

According to the numerical results above, we can use our model to study some 
GRBs whose afterglows 
appear anomalous, such as GRB {000301C}. We employ a standard cosmology with  
$H_0=65{ \rm km~s^{-1}Mpc^{-1}}, 
\Omega_0=1.0$ and $\Lambda=0$. GRB {000301C} is one of the latest afterglows 
exhibiting a sharp break in the optical light curve. 
The redshift was measured using the Hubble space telescope to be $1.95\pm0.1$
by  Smette {et al}.$^{[18]}$ and was later refined by Castro {et al}.$^{[19]}$
using  the Keck 10-m telescope to a value of $2.0335\pm0.0003$, corresponding
to  $D_L\approx13{\rm Gpc}$. Figure 3 shows our numerical fit to the R-band 
light curve of GRB {000301C}. 
We note that the radio afterglow of GRB {000301C} shows a break around 10 days
after the burst,  which is similar to the radio afterglow of GRB {980519}. 
This suggests that the circum-burst environment may be a dense medium.$^{[23]}$
Dai {et al}.$^{[24,25]}$ have studied the dense medium model of GRB afterglows
in detail. Hence we use  our homogeneous model with the following fixed values:
$E_0=1.0\times{10^{51}}{\rm ergs}, M_0=5.0\times10^{-6}{\rm M_{\odot}},
P_i=0.5{\rm ms}, n_{*}=5\times10^4{\rm cm^{-3}}, 
p=3.0, \xi_e=0.02, \xi_B=0.0005, B_s=3.0\times10^{13}{\rm G}$ and $D_L=13{\rm Gpc}$.
The R-band afterglow can be described by 
a power-law with the index $\alpha\sim1.2$ when $1.3<t<7.3{\rm d}$ and
$\alpha \sim 2.7$ when $7.3<t<31.5{\rm d}$, and the break at $t=7.3{\rm d}$
which is well fitted in our model.

In this letter, the significant results of our calculation show that
afterglow light curves may have  two breaks. Then our model should be well
fitted to the light curves of GRB afterglows  exhibiting one break or two
breaks, which depart from the explanation of the standard model.  Recently,
Dai \& Lu$^{[26]}$have used this model to analyze the unusual optical afterglow
of GRB {000301C}. The first break occurs at the time when the energy
injection becomes significant, and this timescale is estimated by $\tau\approx
5\times10^7 E_{0,51}B_{\perp,12}^{-2}P_{i,ms}^4 R_6^{-6}{\rm s}$. When we take
different magnetic field strength ($B_s=10^{12},10^{13}$ and $10^{14}{\rm
G}$), $\tau$ corresponds to the different timescales ($\tau=5\times10^7,
5\times10^5$ and $5\times10^3{\rm s}$) which are consistent with our calculated
results. 

We have numerically compared two types of external medium: homogeneous and wind cases. 
Note that the light curves in the wind model are steeper than those in the homogeneous case, 
so one may use the wind model to explain some afterglows that have decayed rapidly 
at later times which may provide a way to 
distinguish the two cases, giving some hints in understanding GRB progenitor 
environments. However, the sideways-expansion effect of a jet can lead to
similar steep light curves,$^{[13,27-29]}$ so one urgently know how to further
distinguish the wind and jet effects. Now it is  widely argued that the break
may be due to a jet-like outflow at $\gamma\sim\theta^{-1}$, where $\theta$ is
the half opening angle of the jet. 

During our numerical calculations, if we enlarge the value of initial fireball energy,
we find that the light curves have no apparent flattening even we change the 
surface magnetic field strength of a pulsar. We estimate the total energy
injection from the pulsar through magnetic dipole radiation, 
$E_{tot}=2.0\times10^{52}P_{i,ms}^{-2}I_{45}$ in units of erg , which is
independent of the magnetic field strength. When we take  $P_i=1{\rm ms}$ and
$I=10^{45}{\rm g~cm^2}$, $E_{tot}$ is only about  $10^{52}{\rm erg}$, which
is negligible compared to the initial fireball energy up to $10^{53}$ or
larger. Since the moment of inertia is in a narrow range  for different
equations of state (EOS) at high density,  we have to decrease the period of
the pulsar if we need to explain the afterglows of some GRBs  with high
initial energy observed. Note that the submillisecond period for a neutron
star is difficult to achieve using reasonable EOSs  for normal neutron
matter.$^{[30]}$ However, strange stars can reach much shorter periods  than
neutron stars due to the existence of high viscosity in strange stars.$^{[31]}$
If the pulsar with submillisecond period would exist, it may be a strange
star  rather than a neutron star. We fit the R-band light curve of GRB
{000301C} with $P_i=0.5{\rm ms}$,  implying that the pulsar, if it was the
center engine of the fireball, may be a strange star  in our model. Therefore,
some unusual afterglows might provide a method of investigating the nature of
millisecond pulsars if these compact stars are the central engine of GRBs.

\begin{figure}
\plotfiddle{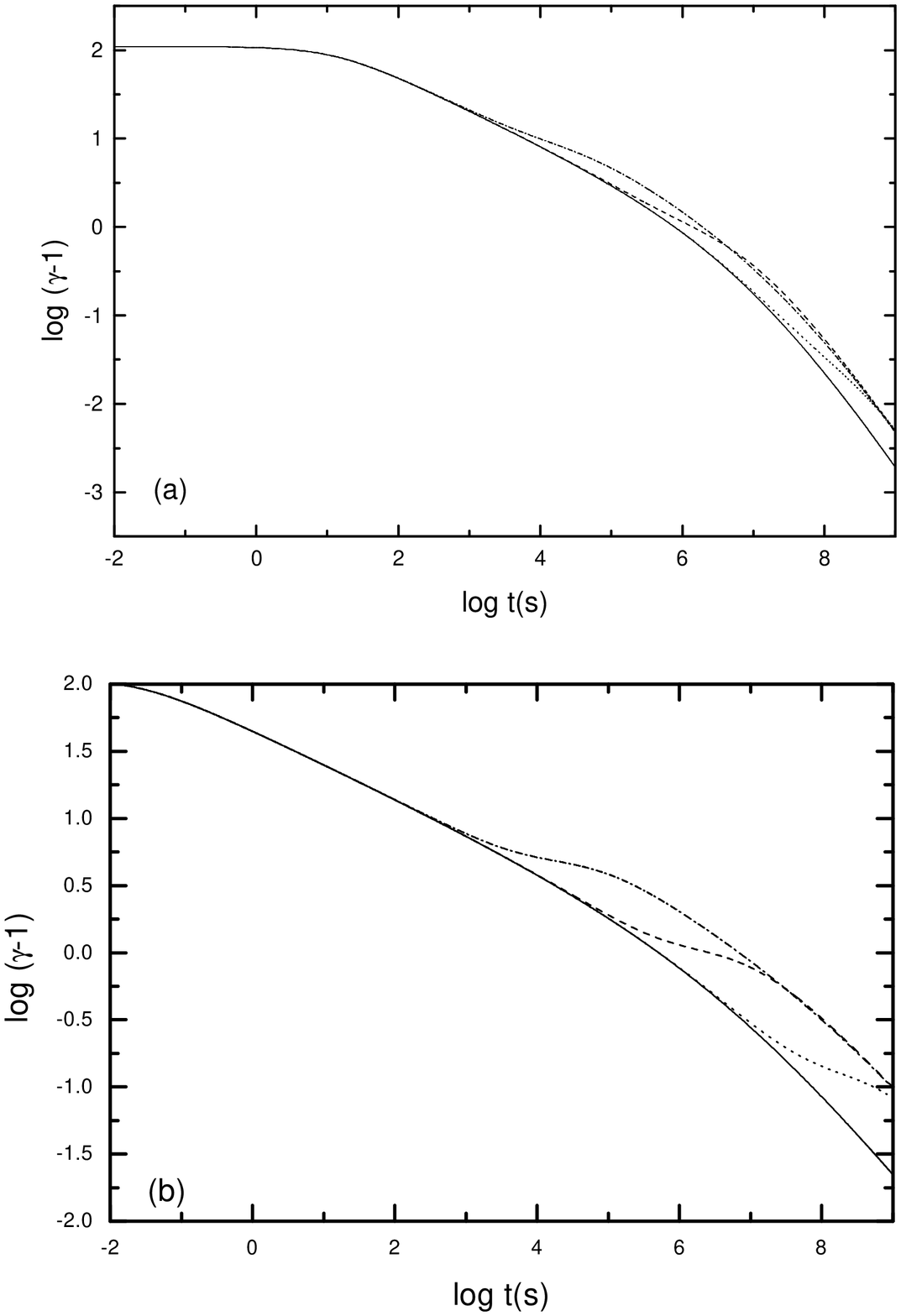}{7in}{0}{60}{70}{-180}{0}
\caption{Evolution of the fireball's Lorentz factor in homogeneous ISM
model(a) and wind model(b). The solid line is plotted with the model without
energy injection, the dotted line corresponding to $B_s=10^{12}{\rm G}$, the
dashed line to  $B_s=10^{13}{\rm G}$, the dash-dot line to
$B_s=10^{14}{\rm G}$. The other parameters and values are presented in the text.}
\end{figure}
\clearpage

\begin{figure}
\plotfiddle{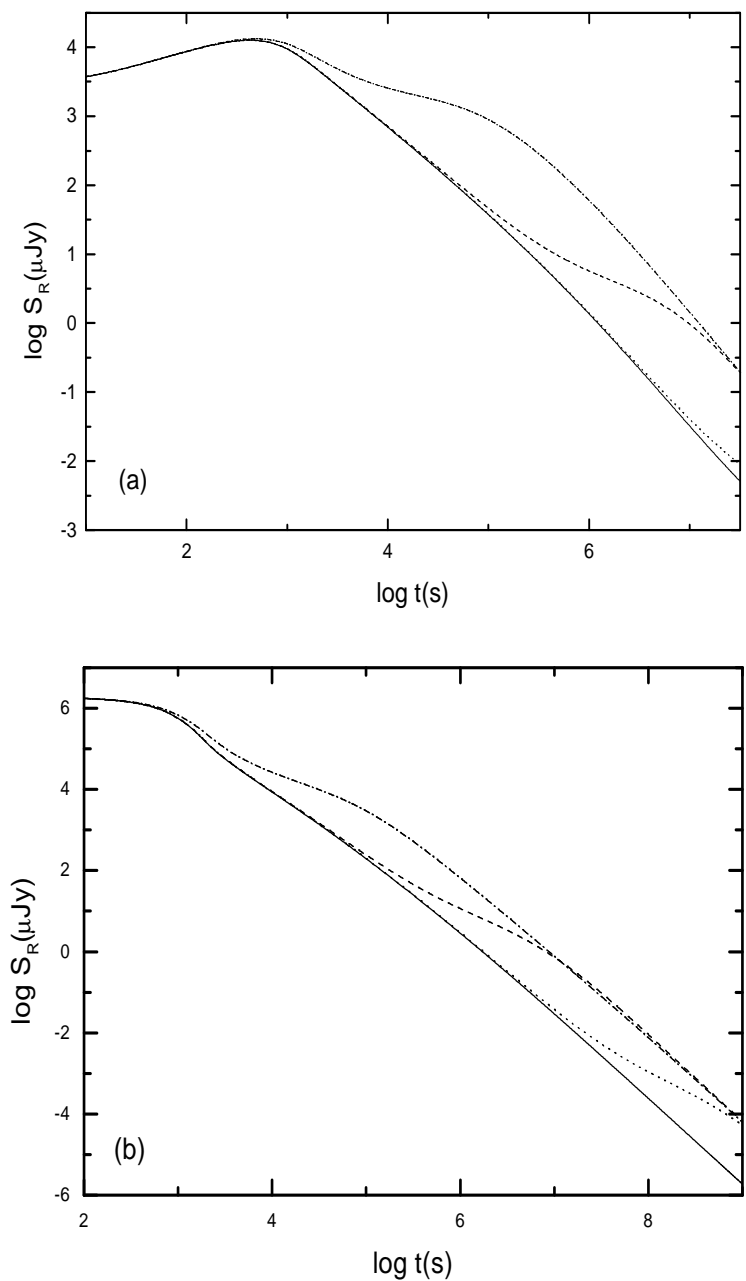}{7in}{0}{60}{70}{-180}{0}
\caption{Predicted optical afterglows in the R-band with two different medium models.
We take $p=2.5, \xi_e=0.1, \xi_B=0.01, D_L=3{\rm Gpc}$.
Other parameters and values are same as in fig.1.}
\end{figure}
\clearpage

\begin{figure}
\plotfiddle{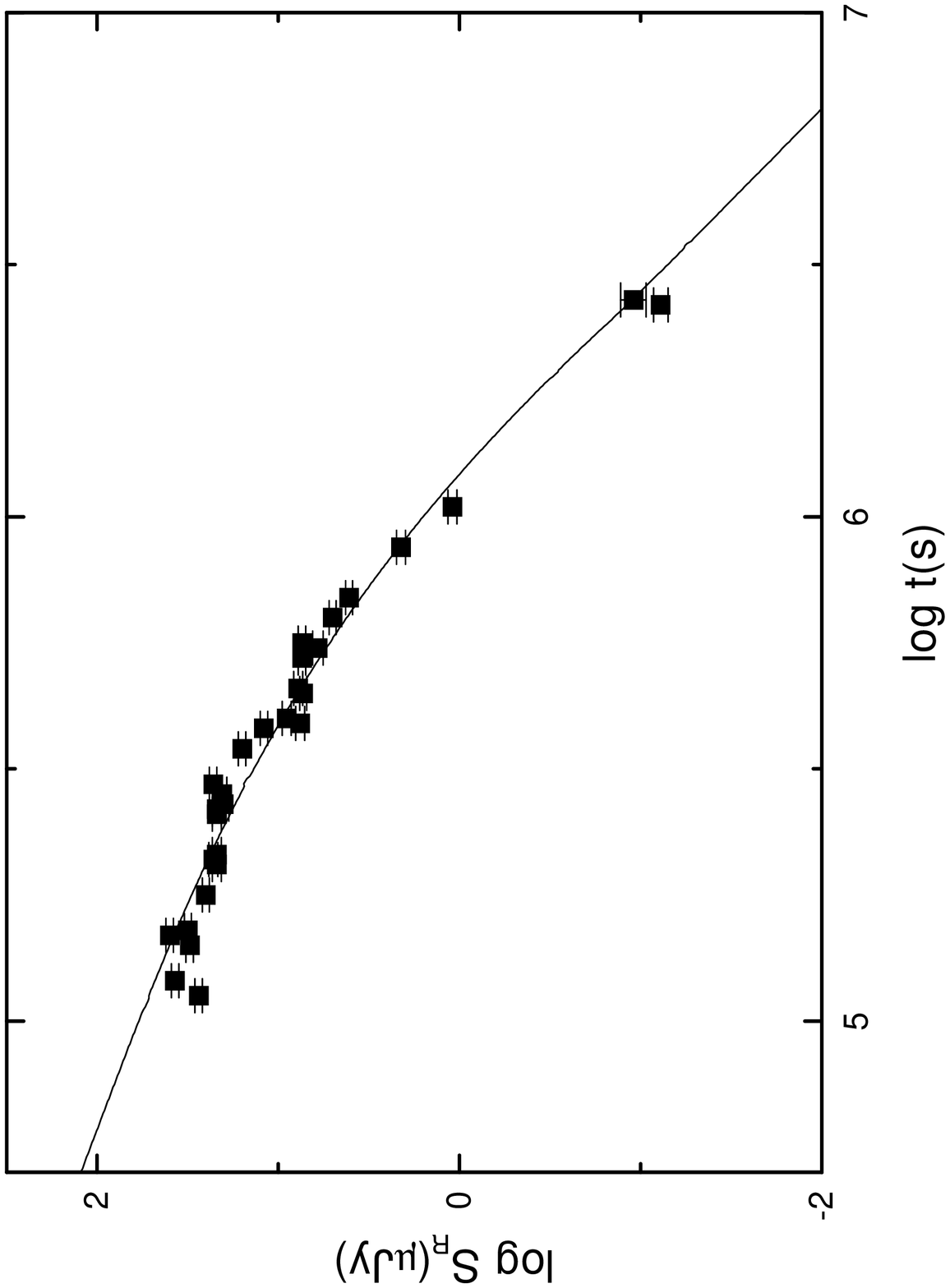}{5.5in}{270}{60}{70}{-250}{400}
\caption{Optical light curve for the R-band afterglow of GRB {000301C}. The
data are taken from Rhoads \& Fruchter$^{[20]}$, Sagar {et al}.$^{[21]}$ and
Masetti {et al}.$^{[22]}$ and we  added 5\% systematic uncertainty. We fit
those data with our homogeneous model: $E_0=1.0\times{10^{51}}{\rm ergs},
M_0=5.0\times10^{-6}{\rm M_{\odot}}, P_i=0.5{\rm ms}, n_{*}=5\times10^4{\rm
cm^{-3}}, \xi_e=0.02, \xi_B=5.0\times10^{-4}, B_s=3.0\times10^{13}{\rm G},
p=3.0$ and $D_L=13{\rm Gpc}$. See the text for more details.} 
\end{figure}

\end{document}